\documentclass[conference]{IEEEtran}
\usepackage{cite}
\usepackage{amsmath,amssymb,amsfonts}
\usepackage{algorithm}
\usepackage{algpseudocode}
\usepackage{graphicx}
\usepackage{textcomp}
\usepackage{xcolor}
\def\BibTeX{{\rm B\kern-.05em{\sc i\kern-.025em b}\kern-.08em
    T\kern-.1667em\lower.7ex\hbox{E}\kern-.125emX}}

\DeclareMathAlphabet{\mcal}{OMS}{cmsy}{m}{n}
\SetMathAlphabet{\mcal}{bold}{OMS}{cmsy}{b}{n} % using a custom name for default mathcal font

\begin{document}

\title{Real-time enforcement of local energy market transactions respecting distribution grid constraints}

\author{\IEEEauthorblockN{Jos\'e Horta}
\IEEEauthorblockA{\textit{Telecom Paristech} \\
Paris, France \\
jose.horta@telecom-paristech.fr}\\
\IEEEauthorblockN{Daniel Kofman}
\IEEEauthorblockA{\textit{Telecom Paristech}\\
Paris, France \\
daniel.kofman@telecom-paristech.fr}
\and
\IEEEauthorblockN{Eitan Altman}
\IEEEauthorblockA{\textit{INRIA}\\
Sophia-Antipolis, France \\
eitan.altman@sophia.inria.fr}\\
\IEEEauthorblockN{David Menga}
\IEEEauthorblockA{\textit{EDF R\&D}\\
Palaiseau, France \\
david.menga@edf.fr}
\and
\IEEEauthorblockN{Mathieu Caujolle}
\IEEEauthorblockA{\textit{EDF R\&D}\\
Palaiseau, France \\
mathieu.caujolle@edf.fr}
}

%\author{\IEEEauthorblockN{Jos\'e Horta\IEEEauthorrefmark{1}, Eitan Altman\IEEEauthorrefmark{2}, Mathieu Caujolle\IEEEauthorrefmark{3}, Daniel Kofman\IEEEauthorrefmark{1}, David Menga\IEEEauthorrefmark{3}}\\
%	\IEEEauthorblockA{\IEEEauthorrefmark{1}Telecom Paristech, Paris, France\\ Emails: \{jose.horta, daniel.kofman\}@telecom-paristech.fr}
%	\IEEEauthorblockA{\IEEEauthorrefmark{2}INRIA, Sophia-Antipolis, France\\ Email: eitan.altman@sophia.inria.fr}
%	\IEEEauthorblockA{\IEEEauthorrefmark{3}EDF R\&D, Palaiseau, France\\ Emails: \{mathieu.caujolle, david.menga\}@edf.fr}}

\maketitle

\begin{abstract}
Future electricity distribution grids will host a considerable share of the renewable energy sources needed for enforcing the energy transition. Demand side management mechanisms play a key role in the integration of such renewable energy resources by exploiting the flexibility of elastic loads, generation or electricity storage technologies. In particular, local energy markets enable households to exchange energy with each other while increasing the amount of renewable energy that is consumed locally. Nevertheless, as most ex-ante mechanisms, local market schedules rely on hour-ahead forecasts whose accuracy may be low. In this paper we cope with forecast errors by proposing a game theory approach to model the interactions among prosumers and distribution system operators for the control of electricity flows in real-time. The presented game has an aggregative equilibrium which can be attained in a semi-distributed manner, driving prosumers towards a final exchange of energy with the grid that benefits both households and operators, favoring the enforcement of prosumers' local market commitments while respecting the constraints defined by the operator. The proposed mechanism requires only one-to-all broadcast of price signals, which do not depend either on the amount of players or their local objective function and constraints, making the approach highly scalable. Its impact on distribution grid quality of supply was evaluated through load flow analysis and realistic load profiles, demonstrating the capacity of the mechanism ensure that voltage deviation and thermal limit constraints are respected.  
\end{abstract}

\section{Introduction}
We consider future distribution grids in a context of massive deployment of Renewable Energy Sources (RES) and storage technologies. Maintaining quality of supply in such a context will require infrastructure reinforcements as usual, but also an active participation of Distribution System Operators (DSO) into the energy management of flexible Distributed Energy Resources\footnote{Flexible loads, controllable generation and storage resources.} (DER). This can be achieved through Demand Side Management, in particular, by gaining access to flexibility from residential prosumers. A myriad of demand side management mechanisms have been proposed and applied for the scheduling of such flexible resources, from which the most relevant ones are those that provide incentives to prosumers for their active participation in distribution grid energy management as selfish agents \cite{Molderink,Samadi,Mohsenian-Rad2010,Li, horta2017}. In particular, we will consider local renewable energy markets that enable households to agree with each other one hour-ahead of time on the exchange of renewable energy. Such markets, as well as all energy management mechanisms that rely on ex-ante agreements (day-ahead, hour-ahead, etc.), imply agents subscribing commitments on future energy flows, as a result of market transactions. Nevertheless, such flows cannot be forecasted with high precision due to their dependency on human behavior and on exogenous variables controlled by nature (temperature, sunshine and wind). Thus, in order to cope with the uncertainty of demand and local renewable production, a real-time control of energy resources must provide the means to drive households towards a final exchange with the grid that benefits both prosumers and DSOs.

In this work we address the control of electricity flows on a residential low voltage distribution grid in real-time. We propose a game theory model that enables households to decide which electricity flows to exchange with the distribution grid in order to minimize their costs. Such costs depend on the average strategy of all households and also on a penalty signal associated to a coupling constraint. This aggregative game enables households to optimize their own costs while reducing voltage rises/drops and current flows on the lines.

The main contributions of this work are the following:
\begin{itemize}
	\item We rely on game theory to model the interactions between a DSO and prosumers for the control of their electricity flows in real-time.
	\item The proposed game has a competitive aggregative equilibrium that can be attained through dynamics that only require one-to-all broadcast of price signals that do not depend on the amount of players.
	\item To the best of our knowledge our work is the first to propose a real-time mechanism for the control of residential electricity flows in a context of massive deployment of renewable energy sources and storage, capable of enforcing both voltage deviations and thermal limit constraints.
\end{itemize}

The article is structured as follows. We start by presenting related work in Section \ref{sec:relWork}. Then we introduce the system under study in Section \ref{sec:SystemGame}. The model and the aggregative game are described in Sections \ref{sec:aheadModel} and \ref{sec:aggregativeGame} respectively. In Section \ref{sec:SimulResultsRTc} we present the simulations and results. Finally, in Section \ref{sec:Conclusions} we conclude the article and provide some perspectives.  

\section{Related Work}
\label{sec:relWork}
We focus on the literature related to real-time control of electricity flows on distribution grids. Several centralized approaches have been proposed for real-time energy management in distribution grid systems \cite{Cecati,Shi, Borghetti}. The difficulty in such approaches is that flexible DER are owned and controlled by households, which hold the local information needed for a centralized control and will release part of this information only if adequate incentives are provided. In \cite{Vytelingum} and \cite{Ramchurn} authors propose decentralized mechanisms with the objective to flatten the aggregated demand of a large set of households. In \cite{Li} they consider the goal of the electricity supplier is to maximize social welfare, which is achieved in a distributed fashion by households optimizing their own benefits.

Game theory has been applied mainly to day-ahead energy scheduling rather than to real-time control, and particularly to modeling the interactions between the electricity supplier and its clients, with the goal of minimizing energy costs rather than enforcing DSO constraints \cite{Mohsenian-Rad,Mohsenian-Rad2010,Soliman,Li, Deng}. In \cite{Mohsenian-Rad2010} they propose a Stackelberg game model that allows electricity suppliers to define prices leading to an equilibrium that minimizes Peak to Average Ratio. In \cite{Soliman} they follow the same approach from \cite{Mohsenian-Rad2010}, but with a strictly convex cost function. In \cite{Deng}, an online version of a scheduling mechanism for flexible appliances is proposed, which copes with price prediction errors. The literature related to real-time control of electricity flows on distribution grids does not consider RES and storage resources or does not take particular care of voltage deviations or thermal constraints. 

%In regards to the state of the art described above we propose a game model for the control of electricity flows exchanged with the grid in real-time by agents representing households equipped with renewable energy sources and flexible demand or storage. Our real-time model provides realistic conditions in which the results of Grammatico \cite{grammatico2017dynamic} can be applied to show that the game has an aggregative equilibrium, with two main advantages, we rely on real exchanges of energy as a measure of the response of the agents to the control signal, which exploits one-to-all coordination and favors truthful responses from agents.

\section{System Description}
\label{sec:SystemGame}
The low voltage distribution grid under study has one or more three-phase four-conductors feeders to which households (with different distributions of resources: PV, batteries and flexible loads) are connected, all equipped with smart meters and with the possibility of controlling the flows they exchange with the electricity grid thanks to smart inverters \cite{pignier2013energy}. Households agree on the exchange of energy with each other through a local energy market described in Section \ref{subsec:lMarket}. Then, a Real-Time Control mechanism (RTC) described in Section \ref{subsec:RTcontrol} copes with forecast errors by driving households towards a final exchange with the grid that benefits the prosumer and respects the DSO's quality of supply requirements. The RTC mechanism is combined with Dynamic Phase Switching (DPS) as described in Section \ref{subsec:phaseSwitching} for balancing load across phases.

\subsection{Local renewable energy market}
\label{subsec:lMarket}
We consider the scheduling mechanism proposed in \cite{horta2017}, where houses agree one hour-ahead of time, through a local market implemented over a blockchain-based transactive platform, on the amount of energy they exchange with each other on a 10 minutes basis. In addition to the exchange with neighbors, any remaining energy flow is contracted with the supplier. One hour ahead of every time slot, each house fixes the schedule of battery usage until the end of the day (moving window), imposing cycling and depth of discharge constraints aimed to extend the battery lifespan. The DSO provides agents with rewards for locally trading renewable energy when is best for the grid, which are attributed \textit{ex post} to incentive agents to enforce their contracted quantities in real-time by absorbing all or part of their forecast errors with their flexibility budget\footnote{Flexibility from batteries/loads reserved for absorbing forecast errors.}.

\subsection{Real-time control mechanism}
\label{subsec:RTcontrol}
Just before each time slot begins, prosumers update their forecast for the next 10 minutes and decide up to which point the error will be absorbed by their flexibility budget or by the electricity supplier. Deviating from the optimal flexibility schedule has an associated cost due to battery (or comfort) degradation, while deviating from the hour-ahead market committed quantities implies loosing part of the reward as well as the beneficial price. The costs of relying on extra flexibility, the final electricity price and the allocation of hour-ahead market rewards are described in Sections \ref{subsec:batteries}, \ref{subsec:finalPrice} and \ref{subsec:rewards} respectively. In addition, during the time slots of high excess of production or demand, households will be charged a penalty fixed by the DSO through the transactive platform, associated to a constraint on the average flow that neighbors exchange with the grid. Such a price signal allows the DSO to influence the decision of households towards a final exchange with the grid that reduces voltage rises/drops and current flows. %The DSO broadcasts the price signal on the blockchain platform, exploiting its transparency and auditability.

\subsection{Dynamic phase allocation}
\label{subsec:phaseSwitching}
We need to ensure the flows across phases are as evenly distributed as possible in order to improve RTC mechanism performance, which is not aimed at balancing load across phases. We assume the operator is capable of deploying solid state switches in a subset of households and of controlling dynamically their allocation to phases as described in \cite{horta2018}. This functionality is complementary to the RTC mechanism.

\section{Model}
\label{sec:aheadModel}
We consider a set of competitive agents $\mcal{H} = \{1,2,\dots,H\}$ representing prosumers and a set $\mcal{N} = \{1,2,\dots,N\}$ of logical nodes on the feeder where the connections of several households are aggregated\footnote{e.g.: start of a branch or where points of measurement are deployed.}, referred to as points of common coupling (PCC). A household is connected to only one PCC. Households control the amount of energy exchanged with the grid by adapting the demand from batteries or flexible loads\footnote{We model the flexibility of loads as a small battery. A comprehensive model of demand elasticity  %(appliance type, usage pattern, etc.)
is out of the scope of this work.}. The goal of the operator is to influence households' decisions in order to reduce voltage deviations and current flows.

\subsection{State of agents}
The state of agents is composed by demand and production forecasts, by the quantities committed on the hour-ahead local market, and by the state of the battery at the beginning of the current time slot, determined by the State of Charge (SoC) and by the accumulated charge/discharge quantities. For each agent $i \in \mcal{H}$, the ex-ante mechanism scheduled the usage of the battery for the current time slot to be $\hat{s}^i$, in order to exchange $\hat{x}^i = \hat{x}^i_m + \hat{x}^i_u$ with the grid, where $\hat{x}^i_m$ and $\hat{x}^i_u$ are the amounts traded with neighbors and with the supplier, at prices $P_m$ and $P_u$ ($P_f$ in the case of an injection) respectively, under a forecasted gap $\hat{r}^i$ between production %$\hat{g}^i \in \mathbb{R}^N_{\ge 0}$ and consumption $\hat{l}^i \in \mathbb{R}^N_{\ge 0}$ according to the following balance equation
$\hat{g}^i$ and consumption $\hat{l}^i$ according to the following balance equation

\begin{align}
\label{eq:balance}
\hat{x}^i_m + \hat{x}^i_u + \hat{s}^i = \hat{l}^i - \hat{g}^i = \hat{r}^i.
\end{align}

A closer to real-time forecast or a Very Short Term Load Forecast \cite{Hsiao} enables each house to update its gap between consumption and production as $r^i = \hat{r}^i + \tilde{r}^i$, where $\tilde{r}^i = \tilde{l}^i - \tilde{g}^i$ is the error on hour-ahead production and load forecasts.

\subsection{Agents' strategies}
The real-time control of each household aims to decide at which extent the error $\tilde{r}^i$ will be absorbed by the grid or by demand flexibility. Thus, each agent $i$ decides on a strategy $(x^i, s^i)$, where $s^i \in \mathbb{R}^N$ represents the amount of energy charged (negative) or discharged (positive) to/from the batteries and $x^i \in \mathbb{R}^N$ is the total amount of energy to be exchanged with the grid on the current time slot\footnote{All but one of the vector components are equal to zero. In future work this will not be the case, due to electric vehicles that can connect to any PCC.} (includes the energy exchanged with neighbors and with the electricity supplier). The relationship between the battery flow $s^i$ and the grid flow $x^i$ is $x^i + s^i = l^i - g^i = r^i$, the same as in (\ref{eq:balance}).

\subsection{Batteries}
\label{subsec:batteries}
The amount of energy stored in the battery of household $i$ at the end of current time slot is represented by $e^i \in \mathbb{R}^N$. Its evolution in time is given by $e^i = e^i_0 - s^i$, where $e^i_0$ represents the energy on the battery at the beginning of current time slot.

Flexibility operations must respect the following constraint:

\begin{align}
\label{eq:RTbattConst}
\operatorname{max}[e^i_0 - \overline{E}^i,\underline{s}^i, \underline{e}^i] \leq s^i \leq \operatorname{min}[e^i_0 - \underline{E}^i,\overline{s}^i, \overline{e}^i]
\end{align}

bounding battery usage to the intersection of three segments:

\paragraph{Energy capacity} the battery cannot be discharged or charged beyond its minimum or maximum energy levels, $\underline{E}^i$ and $\overline{E}^i$, which usually correspond to a SoC around 10-20\% and 80-90\% respectively, in order to preserve its lifetime. The scheduling mechanism limits the energy capacity available for hour-ahead trading to a range of SoC between 25\% and 75\%, reserving a share of the capacity for the RTC mechanism, which can use a depth of discharge between 10\% and 90\%. 

\paragraph{Power capacity} maximum energy that can be discharged/charged on a single time slot, $\underline{s}^i$ and $\overline{s}^i$ respectively.

\paragraph{Cycling capacity} $\underline{e}^i$ and $\overline{e}^i$ are the charging and discharging cycling constraints. For instance, the hour-ahead scheduling mechanism limits battery usage to one daily cycle, while up to an extra cycle per day can be exploited by the RTC mechanism. This extra capacity is allocated to individual time slots in proportion to the corresponding day-ahead forecasted production, in order to adapt the flexibility available during sun hours when forecast error variability is higher.

%We denote $\mcal{S}^i \in \mathbb{R}^N$ the set of flows that satisfy the constraints described above. 
For batteries being operated myopically in real-time under such constraints, the degradation costs can be expressed as 

\begin{align}
\label{eq:degradation}
C^i_e(s^i) = a^i(s^i-\hat{s}^i)^T(s^i-\hat{s}^i) + {b^i}^T(s^i-\hat{s}^i)	
\end{align}

where $s^i$ and $C^i_e$ represent the use of demand flexibility and its corresponding costs, regardless of the source of flexibility being the battery or the elasticity of demand. In the case of batteries, $a^i = a^i_s$ and $b^i = b^i_s$ are parameters that must be chosen adequately to ensure that the costs for an extra use of the battery are compensated by the economic incentives, depending on battery characteristics \cite{ma2015distributed} and on the flexibility budget reserved for the RTC mechanism. For the case of flexible loads, $a^i = a^i_l$ and $b^i = 0$ represent a purely quadratic disutility function as was already proposed in \cite{Samadi2010, aggregators}.

\subsection{Final electricity prices}
\label{subsec:finalPrice}
With respect to prices, we need to take into account the two components of the hour-ahead commitment $\hat{x}^i$, which are the quantity $\hat{x}^i_m$ traded with neighbors at price $P^i_m$ and the quantity $\hat{x}^i_u$ contracted with the supplier at price $P_u$ (or at Feed-In Tariff (FIT) $P_f$ if the prosumer is selling). The price to pay for the effectively exchanged flow $x^i$ will be the following:

\begin{align}
\label{eq:priceCost}
&C^i_p(x^i) = (x^i-\hat{x}^i_m)p + \hat{x}^i_mP^i_m\\
&\textrm{where} \nonumber\\
&p = \left\{
\begin{array}{ll}
P_u,\\
P_f,
\end{array}
\right.&&\begin{array}{ll}
\textrm{if } x^i >= \hat{x}^i_m,&\\
\textrm{else.}& 
\end{array}\nonumber 
\end{align}

Such a price allocation models the fact that a household $i$ that consumes or injects more than what was agreed on the market is considered to have enforced the market transaction; the excess of consumption or injection will be charged or payed at the price $P_u$ or $P_f$ respectively. Otherwise, a house that consumes\footnote{The logic is similar if an agent agrees to inject energy for their neighbors.} less than agreed on the market will pay $x^i_mP^i_m$, as if she had consumed the committed quantity, but will receive $(x^i_m - x^i)P_f$, which corresponds to automatically selling the difference to the supplier at a lower price ($P_f < P_m < P_u$).

\subsection{Reward allocation}
\label{subsec:rewards}
The reward attribution depends on the deviation from the hour-ahead committed quantities as follows:

\begin{align}
\label{eq:reward}
C^i_r(x^i) = a^i_r(x^i-\hat{x}^i)^T(x^i-\hat{x}^i) - R	
\end{align}

where the value $R$ is defined one hour-ahead by the DSO, for instance as a percentage of the transacted quantity. The reward gets reduced quadratically as the agent $i$ deviates from its hour-ahead committed quantity $\hat{x}^i$. Note that if the deviation is too big the reward becomes a penalty.

\subsection{Aggregative Constraint}
The strategies of households are coupled by an aggregative constraint that enables DSO to indirectly enforce voltage deviations and thermal constraints. The constraint over the average strategy is as follows:
\begin{align}
\label{eq:RTcouplConst}
\underline{c} \leq \textstyle\frac{1}{H}\sum^H_{i=1}x^i \leq \overline{c}
\end{align}
where $\underline{c}$ and $\overline{c} \in \mathbb{R}^N$ are the maximum permitted aggregated flows on each PCC in order to conservatively enforce voltage deviations and thermal constraints. We denote as $\mcal{C} \subset \mathbb{R}^N$, the set of $\frac{1}{H}\sum^H_{i=1}x^i$ such that (\ref{eq:RTcouplConst}) is satisfied.

The constraints are obtained by the DSO through load flow sensibility analysis \cite{Borghetti}. For instance, for a period of high injection we progressively increase the aggregate injection and we allocate the increase to individual household in proportion to their injection on the previous time slot. When a constraint is detected, the injections on the previous iteration are considered as the maximum power transit supported by the grid for the time slot. This procedure requires specific knowledge about the grid infrastructure (nodes, lines, impedances, etc.), which is not necessarily available for distribution grids. Such information can be estimated by using different measurements of power injection, voltage and/or current variations \cite{Zhou2008simplified, Christakou,DekaGraphLearn,DekaTerminal,DekaNoData}. This would require short measurement campaigns after which a model of the grid can be obtained. Machine learning techniques can be applied to detect when a new measurement campaign for a model update is necessary.

\subsection{Problem definition}
\label{subsec:RTgame}
Each agent $i$'s goal is to find a strategy $(x^i, s^i)$ such that
\begin{align}
\label{eq:RTgame}
& x^i \in \operatornamewithlimits{arg\,min}\limits_y J^i(y,u)\\
%J^i\left(y,\textstyle\frac{1}{H}\left(\textstyle\sum^H_{j=1}x^j\right),\lambda\right)\\
&\textrm{s.t.} \nonumber\\
&(\ref{eq:balance}), \ (\ref{eq:RTbattConst}),\nonumber\\
&y \in [-\overline{x}^i,\overline{x}^i] \label{const:maxPow} 
\end{align}

where, $J^i(y,u)$ is the local cost function that depends on the individual flow $y$ and on $u(\textstyle\frac{1}{H}\textstyle\sum^H_{j=1}x^j,\lambda)$, a control (price\footnote{negative in case of peak of production and positive in a peak of demand.}) signal broadcasted (on the transactive platform) by the DSO, which depends on the average strategy flow of houses and on $\lambda$, a penalty associated to the coupling constraint in (\ref{eq:RTcouplConst}). The parameter $\overline{x}^i$ in constraint (\ref{const:maxPow}) corresponds to the maximum power capacity contracted with the supplier by household $i$. We denote with $\mcal{X}^i$ the set of feasible values of $x^i$. 

The cost function $J^i : \mathbb{R}^N \times \mathbb{R}^N \rightarrow \mathbb{R} \cup {\infty}$ is defined as:

\begin{align}
J^i(x^i,u) = f^i(x^i) + u^Tx^i
\end{align}

where the function $f^i(x^i) = C^i_e(x^i) + C^i_r(x^i) + C^i_p(x^i)$ reflects the costs that depend only on the individual strategy flow, which are the sum of battery/comfort degradation costs, reward allocation and final electricity pricing\footnote{Note that $C^i_p(x^i)$ are convex piecewise linear functions, while functions $f^i$ are strictly convex and l-strongly convex for any $l \in \left[0,2(a^i + a^i_r)\right)$.}.

\section{Competitive aggregative game}
\label{sec:aggregativeGame}
The problem described above forms a competitive aggregative game, as the optimal response of an agent depends on the aggregate response of the rest of players and they all share a common penalty associated to a coupling constraint. Grammatico proved in \cite{grammatico2017dynamic} that such a game has an aggregative equilibrium under assumptions on functions ${\{f^i\}}^H_{i=1}$ being l-strongly convex and on compactness, convexity and Slater's qualification \cite{Boyd} of the sets ${\{\mcal{X}^i\}}^H_{i=1}$ and $\mcal{C}$. Note that we need $\mcal{C} \subseteq \frac{1}{H}\sum^H_{i=1}\mcal{X}^i$ to avoid the operator fixing constraints that are not attainable with the feasible responses of households. For this we assume that the DSO can estimate the set $\frac{1}{H}\sum^H_{i=1}\mcal{X}^i$ or that, for instance, households could be asked to communicate their flexibility budget for every time slot in which the real-time control mechanism is activated.
Aggregative equilibrium is defined as follows.

\paragraph*{Definition 1:} \textit{Aggregative equilibrium}.
\label{def:equilibrium}
A tuple $((x^{i\ast})^H_{i=1},\lambda^\ast)$ is an Aggregative equilibrium, for the game described in (\ref{eq:RTgame}) with the coupling constraint in (\ref{eq:RTcouplConst}) if $\frac{1}{H}\sum^H_{i=1}x^{i\ast} \in \mcal{C}$, and for all $i \in \mcal{H}$, 

\begin{align}
\label{eq:aggEquilibrium}
x^{i\ast} \in \operatornamewithlimits{arg\,min}\limits_{y \in \mcal{X}^i} f^i(y) + \left(D\textstyle\frac{1}{H}\left(y + \textstyle\sum^H_{j \neq i}x^{j\ast}\right) + \lambda^\ast\right)^Ty.
\end{align}

\subsection{Iterative Process}
Such an equilibrium can be attained by following the semi-distributed approach proposed by Grammatico \cite[Section III]{grammatico2017dynamic}. The mechanism relies on households responding optimally to the incentive signal $u$. The iterative process to attain an aggregative equilibrium starts at the beginning of the 10 minutes time slot with a period of around 10 seconds\footnote{While the time between blocks on the blockchain-based transactive platform described in \cite{horta2017} is of around 5 seconds.}, which should enable households to adapt to penalty changes, while providing enough iterations for the semi-distributed algorithm to converge to the equilibrium. At the beginning of $k$-th iteration, the operator will estimate the aggregated optimal response $\mcal{A}(u_{(k-1)})$ on the previous iteration by measuring the flows on each PCC of the corresponding feeder. Then it will update the signal $u_{(k)} = \kappa(t, u_{(k-1)})$, using the dynamic control law $\kappa$ proposed in \cite[Section III]{grammatico2017dynamic}. The signal will then reach the prosumers through the next published block on the transactive platform. At the detection of a change in signal $u$, each agent will proceed to update its optimal response $x^{i*}_{(k)}$, by updating its battery optimal flow $s^{i*}_{(k)}$ from (\ref{eq:balance}). 

In each iteration, prosumers decide on the energy flow to exchange with the grid on the whole 10 minutes, while relying on forecast done at the beginning of the time slot and on measurements of energy flows and battery usage up to the current iteration. The relationships on the optimal flows for the entire time slot and the flows up to iteration $k$ are as follows 

\begin{subequations}
	\begin{align}
	& x^{i*}_{(k)} = x^i_{0k} + x^i_{kT} \\
	& s^{i*}_{(k)} = s^i_{0k} + s^i_{kT} \label{eq:kiter}
	\end{align}
\end{subequations}

where $x^i_{0k}$ corresponds to the flow exchanged with the grid up to the $k$-th iteration and $x^i_{kT}$ to the future energy flow up to the end of the time slot. The notation is the same for battery flows. Note that we assume households have access to the cumulated energy exchanged with the grid $x^i_{0k}$, through the smart meter, and to the corresponding battery usage $s^i_{0k}$, through the smart inverter/battery controller. 

The model we apply is deterministic, as it does not consider that the state of agents can change during the negotiation process towards the equilibrium. It relies then on the forecast done by households at the beginning of each time slot to be perfect. An analysis of the  sensibility to very short term forecast errors and the possibility of applying stochastic game theory are subjects of future research.

Under such a deterministic scenario, each household will define the battery setting for the rest of the time slot as follows

\begin{align}
\label{eq:batteryset}
s^i_{kT} = r^i - x^{i*}_{(k)} - s^i_{0k}
\end{align}

where (\ref{eq:balance}) and (\ref{eq:kiter}) were used. Here we assume that households set their batteries to the value $s^i_{kT}$ so that if the flows are maintained up to the end of the time slot the total flow exchanged with the grid during the time slot would be $x^{i*}_{(k)}$.

The iteration procedure towards an aggregative equilibrium is summarized as follows

%\begin{figure}[htb!]
%	\centering
%	\includegraphics[width=\columnwidth]{algorithm.png}
	%\caption{Maximum voltage deviations along the feeder.}
%	\label{algo:determ}
%\end{figure}

\begin{algorithm}
	\caption{Dynamic control of competitive optimal responses \cite{grammatico2017dynamic}}\label{algo:determ}
	\begin{algorithmic}
		\State \textbf{Initialization:}  $t \leftarrow 0$;
		\State $\quad$ $\bullet$ The DSO chooses $u_{(0)}$;
		\vspace{0.3cm}
		\State \textbf{Iterate until convergence:}
		\State $\quad$ $\bullet$ DSO broadcasts $u_{(0)}$ to all agents
		\vspace{0.3cm}
		\State $\quad$ $\circ$ Each agent $i \in \mathcal{H}$ computes $x^{i*}(u_{(t)})$, 
		\State $\quad$ $\circ$ and define battery setting for the rest of the time slot $s^i_{tT}$.
		\vspace{0.3cm}
		\State $\quad$ $\bullet$ DSO measures average best response $\mathcal{A}(u_{(t)})$,
		\State $\quad$ $\bullet$ obtains $u_{(t+1)} = \kappa(t, u_{(t)})$
		\vspace{0.3cm}
		\State $\quad$ $t \leftarrow t+1$
	\end{algorithmic}
\end{algorithm}

\section{Simulations and Results}
\label{sec:SimulResultsRTc}
\subsection{Procedure and system scenario}
The simulation of the real-time control mechanism relies on local energy market results and on an allocation of households to phases updated by a DPS mechanism. Those two additional mechanisms are simulated following the procedures described in \cite{horta2017} and \cite{horta2018} respectively.

We consider the forecast used by households for the hour-ahead market to be a day-ahead persistence forecast, and the updated forecast just before the time slot to be perfect, which allows us to apply a deterministic model.  

With respect to the monitoring period, the RTC mechanism is only launched if during the previous time slot we detected a voltage deviation of 9\% or a current flow above 70\% of the thermal limit. To simulate such monitoring we realize load flow studies for every time slot in high sun hours (from 10 am to 2 pm), and when a risk of constraint is detected the RTC mechanism is launched.

We consider a distribution grid composed of one feeder with 50 households, 60\% of which are equipped with a battery and 80\% with a PV panel. All households participate on the market and are eligible for dynamic phase switching. With respect to the level of aggregation of households on each logical PCC, we consider one PCC per phase of the feeder. This means that all the houses on the same phase are considered to share the same logical PCC. This keeps the model simple and independent of the physical structure of the feeder and the distribution of households. Furthermore, the measurements needed by the DSO to implement the real-time control mechanism are currently available at this level of aggregation without the need of deploying further measurement devices.

\subsection{Simulation tools and parameters}
To implement the mechanism described in Algorithm \ref{algo:determ}, we use MATLAB \cite{MATLAB} together with the Gurobi \cite{Gurobi} optimization suite. For the power flow analysis, we rely on the Distribution Network Simulation Platform (DisNetSimPl) developed by EDF R\&D. We rely on an electricity network model conformed by a 20 kV/410 V transformer of 160 kVA rated power and a feeders with 70 $mm^2$ aluminum power lines.

The parameters used for the simulations were the following:

\paragraph*{Load profiles -}We use realistic synthetic consumption data obtained from the Multi-agent Simulator of Human Behavior SMACH \cite{amouroux2013simulating} as input for the household optimization problem. The load curves correspond to 7 winter days of consumption on a 1-minute basis from 50 households of mixed profiles.
\paragraph*{Production profiles -}We consider the same synthetic production curve for all the PV panels (all located on the same area), but the profile varies on a daily basis. 
\paragraph*{Batteries -}We consider ideal batteries of 9 kWh total capacity. Up to 6 kWh (10\% to 77\%) are allowed to be used for local market exchanges in order to preserve the battery lifespan while reserving a share (1.2 kWh) of the capacity to the RTC mechanism, which can use up to 7.2 kWh (10\% to 90\%) of the capacity and perform an additional cycle\footnote{This would be similar to allowing batteries to cycle up to two times in a day. Nevertheless, we preallocate the flexibility budget to individual time slots proportionally to the forecasted energy production, without the possibility of accumulating the budget that is not used.} for absorbing forecast errors. For the case of flexible appliances, whose elasticity is entirely reserved for the RTC mechanism, we consider up to 1 kWh of energy is available for coping with forecast errors\footnote{A higher amount could be considered taking into account that only a water heater can consume more than 5 kWh a day}.   
\paragraph*{Electricity prices -}We consider a common supplier offering a Time Of Use pricing with two levels: 15 c€/kWh from 12 am to 4 pm and 20 c€/kWh from 5 pm to 11 pm. While the FIT is considered to be 10 c€/kWh. 

\subsection{Results and discussion}
We analyze the performance of the RTC mechanism during the time slots with high renewable energy production (between 10 am and 2 pm). In particular, we are interested in cases where we observe that thermal or voltage limits are breached, or close to the limits, for the scenario without RTC. From the 7 days analyzed on the simulations we focus only on the most critical one. First, we analyze voltage deviations with respect to the nominal value $Un$ and then the reduction of current intensity over the lines. We compare three cases, one without any control from the DSO in real-time, one with DPS only and one applying the RTC mechanism coupled with DPS. For the first two cases households optimize their electricity bill with the signal $u$ being $0$ for all time slots.

\subsubsection{Reduction of Voltage deviations}

\begin{figure}[htb!]
	\centering
	\includegraphics[width=\columnwidth]{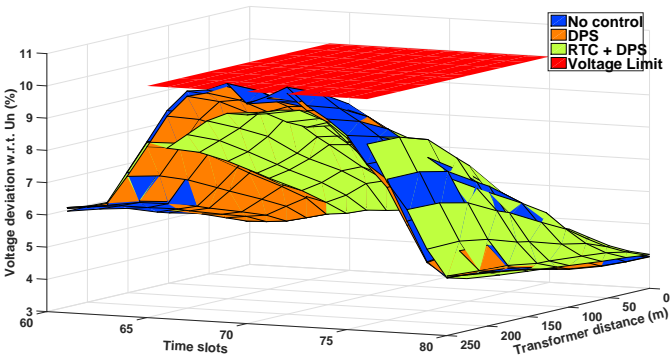}
	\caption{Maximum voltage deviations along the feeder.}
	\label{fig:maxVriseD1F1}
\end{figure}

We start by analyzing the impact of the RTC mechanism in voltage deviations along the feeder. In Figure \ref{fig:maxVriseD1F1}, we see four surfaces, the red (planar) surface shows the voltage limit during the period in which RTC was active (from time slot 64 up to 74), the dark blue one shows the voltage deviations observed without any DSO control, the light blue one with only DPS being applied and the green one applying the RTC mechanism coupled with DPS. While for the cases without RTC we can observe the voltage limits being violated (when the corresponding surfaces go above the limit), the application of RTC combined with DPS achieves consistent reductions of (maximum) voltage deviations. This can be observed as the green surface (RTC + DPS) goes below the others starting from time slot 64 and up to time slot 74. These results demonstrate the capacity of RTC combined with DPS for avoiding voltage limit breaches. 

\subsubsection{Reduction of current flows through the lines}
\begin{figure}[htb!]
	\centering
	\includegraphics[width=\columnwidth]{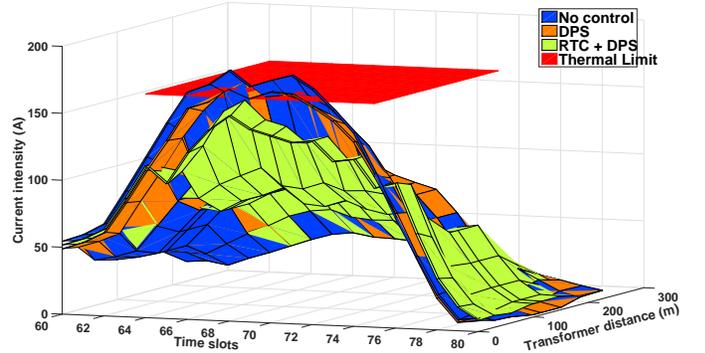}	
	\caption{Current intensity level along the feeder.}
	\label{fig:Ic_D1_half}
\end{figure}

With respect to the current flows through the lines, in Figure \ref{fig:Ic_D1_half} we show the current intensity along the Feeder during the monitoring period. The red (planar) surface shows the thermal limit during the period in which RTC was active. During this period we can clearly appreciate considerable reductions of current intensity all along the feeder avoiding the violation of thermal constraints that are observed for the two cases without RTC\footnote{The rebound effect after the end of the RTC period, as in most demand response applications, happens when current and voltage levels are safer.}. Such reductions are enough to avoid or postpone the replacement of entire line segments, which directly translates into a considerable reduction of infrastructure investment. 

\subsubsection{Dynamic control mechanism performance}
With respect to the performance of the dynamic control mechanism to attain the aggregate equilibrium, we observed an improvement of the convergence when combining RTC with DPS. Without the DPS mechanism the mean amount of iterations was over 60 (10 minutes), which would not be applicable to our setting without further tuning. The DPS mechanism brought the mean convergence time to 5 minutes, with a maximum of 39 iterations, which is adequate for our setting. This is probably due to the fact that our level of logical PCC aggregation is the phase, and if the loads and their DER are better distributed among phases the mechanism converges faster. For instance, if one of the phases has little or no flexibility budget, then the convergence will be slower. After convergence, each household just needs to put the battery to track the optimal flow to exchange with the grid during the rest of the time slot.

\section{Conclusions and perspectives}
\label{sec:Conclusions}

Local energy markets enable households to exchange energy with each other while increasing the amount of renewable energy that is consumed locally. Such markets increase the capacity of distribution grids for hosting renewable energies, but, as all ex-ante mechanisms, local market schedules rely on hour-ahead forecasts whose accuracy may be low. We propose a game-based real-time control mechanism to cope with forecast errors by driving households to a final exchange with the grid that benefits both the prosumer and the DSO. The proposed game has an aggregative equilibrium which can be attained in a semi-distributed manner with a number of iterations independent of the amount of households.

The performance of the mechanism is evaluated through load flow analysis and realistic load curves for a scenario with 50 households, where 80\% of them are equipped with PV panels and 60\% with storage. The simulations show that without the control mechanism the grid would not be capable of hosting such level of penetration of renewable energies. However, when the real-time control mechanism is applied in combination with dynamic phase allocation, maximum voltage deviations and current intensities are considerably reduced, avoiding the violations of voltage deviation and thermal constraints. We show that the proposed real-time mechanism is capable of adjusting the flows issued from local market commitments towards an equilibrium that optimizes the prosumers' electricity bill while increasing the capability of the distribution grid to support the energy transition. 

The performance of the proposed real-time market mechanism can be further improved by better selecting the parameters of the dynamic control scheme and of the local cost functions. The economic performance for the DSO and the performance of the equilibrium with respect to social welfare require further analysis. With respect to the combination with the dynamic phase switching approach, the selection of households and the switching mechanism could be specifically adapted to further reduce voltage deviations. We are currently working on a sensibility analysis to very short term load/production forecast errors that could provide valuable insights for the development of stochastic game models.

\bibliographystyle{IEEEtran}
\bibliography{IEEEabrv,references}

\end{document}